\documentclass[aps,prd,superscriptaddress]{revtex4-2}
\usepackage{amsmath,amssymb,amsfonts,mathrsfs,latexsym,amscd,bm,bbm} 
\usepackage{float}
\usepackage{graphicx,xcolor} 
\usepackage[colorlinks=true,linkcolor=blue]{hyperref}
\usepackage{enumerate} 
\usepackage{siunitx}[=v2]
\usepackage{braket}
\usepackage{slashed} 
\usepackage{makeidx} 
\usepackage{hhline,multirow}
\usepackage{mathtools}

\DeclareMathOperator{\tr}{Tr}
\DeclareMathOperator{\diag}{diag}

\newcommand*{\dif}{\mathop{}\!\mathrm{d}}

\begin{document}

\title{Sea quark contributions to the electromagnetic form factors of $\Sigma$
	hyperons}
\author{Mingyang Yang}
\email{yangmy@ihep.ac.cn}

\affiliation{Institute of High Energy Physics, CAS, P. O. Box 918(4), Beijing 100049,
	China}
\affiliation{College of Physics Sciences, University of Chinese Academy of Sciences,
	Beijing 100049, China}
\author{Ping Wang}
\email{pwang4@ihep.ac.cn}

\affiliation{Institute of High Energy Physics, CAS, P. O. Box 918(4), Beijing 100049, China}

\begin{abstract}
	We study the sea quark contributions to the electromagnetic form factors of $\Sigma$ baryons with nonlocal chiral effective theory. Both octet and decuplet intermediate states are included in the one loop calculation. $G_{\Sigma^{-}}^{u}$ and $G_{\Sigma^{+}}^{d}$ could be priority observables for the examination of sea quark contributions to baryon structure because these quantities are much larger than the strange form factors of nucleon. It will be less difficult for lattice simulation to determine the sign of these pure sea quark contributions unambiguously. In $\Sigma^0$, the light sea quark form factors $G_{\Sigma^{0}}^{u}$ and $G_{\Sigma^{0}}^{d}$ are identical. Since the light sea quark form factors in proton are different, it will be more meaningful to compare lattice result of the light sea quark form factors in $\Sigma^0$ with that obtained from effective field theory. 
	
\end{abstract}
\maketitle

\section{introduction}

Electromagnetic form factors of nucleon are among the most important observables of these building blocks of the ordinary matter in the universe. A lot of theoretical and experimental efforts have been made for many years to get the precise values of the form factors. The form factors' data have triggered much activity in the determinations of the flavor separated form factors of the dressed up, down and strange quarks in the nucleon. It is well known that a complete characterization of the baryon substructure must go beyond three valence quarks. In particular, the strange quark contribution to the nucleon electromagnetic form factors is of special interests because it is purely from the sea quark. The strange form factors of the proton $G_E^s$ and $G_M^s$ have been obtained by a number of groups, such as SAMPLE at Bates \cite{Sample}, A4 at Mainz \cite{A41,A42}, G0 \cite{G01,G02} and HAPPEX \cite{Happex1,Happex2,Happex3,Happex4} at Jefferson Lab, etc. However, up to now, the experiments are not able to determine the signs of $G_E^s$ and $G_M^s$ unambiguously although the global analyses do tend to suggest that $G_M^s < 0$ is favored \cite{Young,Jimenez}.

Theoretically, the strange form factors were investigated in various phenomenological models which give different predictions \cite{Jaffe,Park,Cohen,Musolf,Weigel,Hammer,Hong,Hannelius,Silva,Lyubovitskij,Zou}. The strange form factors were also investigated in the chiral perturbation theory (ChPT) or the effective field theory (EFT) \cite{Musolf2,Hemmert,Kubis}. Historically, most formulations of ChPT are based on the dimensional or the infrared regularization. Though ChPT is a successful and systematic approach, for the nucleon electromagnetic form factors, it is only valid for $Q^2 < \SI{0.1}{\square\GeV}$ \cite {Fuchs}. When vector mesons are included, the results are close to the experiments with $Q^2$ less than $\SI{0.4}{\square\GeV}$ \cite{Kubis2}. Because of the unknown low energy constants appearing in the chiral Lagrangian, the capacity to predict the strange form factors is much limited. In other words, the strangeness vector current matrix elements that one wants to predict, are the same quantities one needs to know in order to make such predictions. \cite{Musolf2,Kubis}.

An alternative regularization method, namely the finite-range-regularization (FRR) has been proposed. 
Inspired by quark models that account for the finite-size of the nucleon, as the source of the pion cloud, effective field theory with FRR has been widely applied to extrapolate the vector meson masses, magnetic moments, magnetic form factors, strange form factors, charge radii, first moments of GPDs, and the nucleon spin, etc \cite{Allton,Wang0,Li}. With the regulators, the loop integrals are convergent. 
One advantage of ChPT with the FRR is that strange form factors can be predicted with the assumption that there is no valence contribution from the tree level.
No low energy constants are needed to cancel the loop divergence, for predicting the strange form factors  \cite{Wang1,Wang2}.

In recent years, we proposed a nonlocal chiral effective theory, which makes it possible to study the form factors at relatively large $Q^2$ \cite{He1,He2,Yang}. 
With the introduction of the gauge links, the nonlocal Lagrangians are locally gauge invariant. The nonlocal interaction generates both the relativistic regulators which make the loop integrals convergent and the $Q^2$ dependence of form factors at tree level. The obtained electromagnetic form factors and strange form factors of nucleon are very close to both experimental data and lattice results \cite{He1,He2}. This nonlocal EFT was further applied to study the parton distribution functions \cite{Salamu1,Salamu2}. 
For the sea quark form factors of proton, the light sea quark contributions obtained in nonlocal EFT are quite different from the lattice results, though the strange quark contributions are comparable with each other \cite{Yang2,Suan,Alexandrou}. 
There is an obvious flavor asymmetry of the $u$--sea and $d$--sea in nonlocal EFT, while the light sea quark contributions are the same for $u$ and $d$ in the lattice simulation.

Therefore, to get a better understanding of the sea quark properties, in this paper, we investigate the sea quark contributions to the electromagnetic form factors of $\Sigma$ hyperons. 
For $\Sigma^+$ ($\Sigma^-$), the $d$ ($u$) quark contribution is purely from the sea, which is the same as the $s$ quark contribution in nucleon. 
However, because the $d$ ($u$) quark contribution in $\Sigma^+$ ($\Sigma^-$) comes from the $\pi$ meson loop, 
it is much larger than the strange quark contribution in nucleon, concerning the latter comes from the $K$ meson loop. 
It will be easier to get an unambiguous number for the lattice simulations, in spite of the statistic errors. 
For the experimentalists, it may also be possible to determine the signs of the light sea quark contributions in $\Sigma$ hyperons, since the magnetic moments of $\Sigma^+$ and $\Sigma^-$ have been measured precisely for many years. 
The investigation of $d$ ($u$) quark contribution to $\Sigma^+$ ($\Sigma^-$) form factors, will definitely shed light on the signs of nucleon strange form factors. 
As pointed in Ref.~\cite{Zou}, the signs of the strange form factors will indicate the possible configuration of $uuds\bar{s}$ components of the proton. 
It could inform us whether the sea quark is in the baryon-meson configuration, or in the diquark configuration.

The flavor asymmetries of light sea quark will be also obvious in $\Sigma^+$ and $\Sigma^-$, as those in proton.
We hope further lattice simulations could distinguish these sea quark asymmetries.
To corroborate the comparison, which between the lattice simulation and the nonlocal EFT, 
we calculate the light sea quark contributions in $\Sigma^0$, 
where the $u$--sea and $d$--sea do have the same contributions. 
In this case, one can make a direct comparison between these two calculations. 
This paper is organized in the following way. 
The nonlocal Lagrangian is introduced in section II. In section III, the sea quark form factors of $\Sigma$ hyperons will be presented. Numerical results will be discussed in section IV. Finally, section V is a short summary.

\section{nonlocal effective Lagrangian}

The lowest order chiral Lagrangian involving baryons, pseudo-scalar mesons and their interactions can be written as \cite{He1,Salamu2,Jenkins1,Jenkins2}
\begin{align}
	\mathcal{L} & =i\tr\bar{B}\left(\gamma_{\mu}\slashed{\mathscr{D}}^{\mu}-m_{B}\right)B+
	\bar{T}_{\mu}^{abc}(i\gamma^{\mu\nu\alpha}D_{\alpha}-m_{T}\, \gamma^{\mu\nu})T_{\nu}^{abc}+
	\frac{f^{2}}{4}\tr\left(\partial_{\mu}U\partial^{\mu}U^{\dagger}\right)\nonumber\\
    & +D\tr\left(\bar{B}\gamma_{\mu}\gamma_{5}\{A^{\mu},B\}\right)+
    F\tr\left(\bar{B}\gamma_{\mu}\gamma_{5}[A^{\mu},B]\right)+
    \frac{\mathcal{C}}{f}\varepsilon^{abc}\bar{T}_{\mu,a}^{\phantom{\mu,a}de}(g^{\mu\nu}+
    z\gamma^{\mu}\gamma^{\nu})B_{ce}\partial_{\nu}\phi_{bd}+H.C,
\end{align}
where $D$, $F$ and ${\cal C}$ are the coupling constants. $z$ is the off-shell parameter. 
The chiral covariant derivative $\mathscr{D}_{\mu}$
is defined as $\mathscr{D}_{\mu}B=\partial_{\mu}B+\left[V_{\mu},B\right]$.
The pseudo-scalar meson octet couples to the baryon field via the vector and axial-vector combinations
\begin{align}
	V_{\mu} & =\frac{1}{2}\left(\zeta\partial_{\mu}\zeta^{\dagger}+\zeta^{\dagger}\partial_{\mu}\zeta\right)+\frac{1}{2}ie\mathscr{A}_{\mu}\left(\zeta^{\dagger}Q_{q}\zeta+\zeta Q_{q}\zeta^{\dagger}\right),\nonumber \\
	A_{\mu} & =\frac{i}{2}\left(\zeta\partial_{\mu}\zeta^{\dagger}-\zeta^{\dagger}\partial_{\mu}\zeta\right)-\frac{1}{2}e\mathscr{A}_{\mu}\left(\zeta Q_{q}\zeta^{\dagger}-\zeta^{\dagger}Q_{q}\zeta\right),
\end{align}
where
\begin{equation}
	\zeta^{2}=U=e^{i2\phi/f},\qquad f=\SI{93}{MeV}.
\end{equation}
$Q_{q}$ is the real-valued matrix of charge, $\diag\left(2/3,-1/3,-1/3\right)$.
$\phi$ and $B$ are the matrices of pseudo-scalar meson fields and octet baryons, respectively.
$\mathscr{A}_{\mu}$ is the photon field. The covariant derivative $D_{\mu}$ in the decuplet sector is defined as $D_{\nu}T_{\mu}^{abc}=\partial_{\nu}T_{\mu}^{abc}+\left(\Gamma_{\nu},T_{\mu}\right)^{abc}$,
where $\Gamma_{\nu}$ is the chiral connection defined as $\left(X,T_{\mu}\right)^{abc}=\left(X\right)_{\phantom{a}d}^{a}T_{\mu}^{dbc}+\left(X\right)_{\phantom{b}d}^{b}T_{\mu}^{adc}+\left(X\right)_{\phantom{c}d}^{c}T_{\mu}^{abd}$.
Besides, $\gamma^{\mu\nu\alpha}$, $\gamma^{\mu\nu}$ are the anti-symmetric matrices expressed as
\begin{equation}
	\gamma^{\mu\nu}=\frac{1}{2}\left[\gamma^{\mu},\gamma^{\nu}\right],\hspace{0.5cm}\hspace{0.5cm}\gamma^{\mu\nu\rho}=\frac{1}{4}\left\{ \left[\gamma^{\mu},\gamma^{\nu}\right],\gamma^{\rho}\right\} .
\end{equation}

The octet, decuplet and octet-decuplet transition operators for magnetic interactions, are needed in the one loop calculations. The anomalous magnetic Lagrangian of octet baryons is written as
\begin{equation}
	\mathcal{L}_{\text{oct}}=\frac{e}{4m_{B}}\left[c_{1}\tr\left(\bar{B}\sigma^{\mu\nu}\left\{ \mathscr{F}_{\mu\nu}^{+},B\right\} \right)+c_{2}\tr\left(\bar{B}\sigma^{\mu\nu}\left[\mathscr{F}_{\mu\nu}^{+},B\right]\right)+c_{3}\tr\left(\bar{B}\sigma^{\mu\nu}B\right)\tr\left(\mathscr{F}_{\mu\nu}^{+}\right)\right],
\end{equation}
where
\begin{equation}
	\mathscr{F}_{\mu\nu}^{+}=-\frac{1}{2}\mathscr{F}_{\mu\nu}\left(\zeta^{\dagger}Q_{q}\zeta+\zeta Q_{q}\zeta^{\dagger}\right).
\end{equation}
The above Lagrangian will contribute to the Pauli form factor $F_{2}^{f,\Sigma}$ which is defined in Eq.\eqref{eq:f1f2}. At the lowest order, the contribution of quark $q$ with unit charge to the octet magnetic moments, can be obtained by replacing the charge matrix $Q_{q}$ with the corresponding diagonal quark matrices $\lambda_{q}=\diag\left(\delta_{qu},\delta_{qd},\delta_{qs}\right)$. After expansion of the above equations, it is found that, taking $\Sigma$ hyperons as an example,
\begin{align}
	 & F_{2}^{u,\Sigma^+}=c_{1}+c_{2}+c_{3},\quad F_{2}^{d,\Sigma^+}=c_{1}-c_{2}+c_{3},\quad F_{2}^{s,\Sigma^+}=c_{3},\nonumber \\
	 & F_{2}^{u,\Sigma^-}=c_{1}-c_{2}+c_{3},\quad F_{2}^{d,\Sigma^-}=c_{1}+c_{2}+c_{3},\quad F_{2}^{s,\Sigma^-}=c_{3}.
\end{align}
Comparing with the results of the constituent quark model, where $F_{2}^{d,\Sigma^+}=0$, and $F_{2}^{u,\Sigma^-}=0$, we get
\begin{equation}
	c_{3}=c_{2}-c_{1}.
\end{equation}
The above relationship is consistent with that there is no strange quark contribution in bare nucleon.

The magnetic moment operators of decuplet and octet-decuplet transition are expressed as
\begin{equation}
	\mathcal{L}_{\text{dec}}=-\frac{ieF_{2}^{T}}{4M_{T}}\bar{T}_{\mu,abc}\sigma_{\rho\lambda}F^{\rho\lambda}T^{\mu,abc}
\end{equation}
and
\begin{equation}
	\mathcal{L}_{\text{trans}}=\frac{ie\mu_{T}}{4m_{B}}F_{\mu\nu}\left(\epsilon^{ijk}Q_{c,il}\bar{B}_{jm}\gamma^{\mu}\gamma_{5}T^{\nu,klm}+\epsilon^{ijk}Q_{c,li}\bar{T}^{\mu,klm}\gamma^{\nu}\gamma_{5}B_{mj}\right),
\end{equation}
respectively.

The gauge invariant nonlocal Lagrangian can be obtained using the method in \cite{pingw.quantization,He1,He2}. For instance, the local interaction between $\Sigma$ hyperons and $\pi^{+}$ meson can be written as
\begin{equation}
	\mathcal{L}_{\Sigma\pi^+}^{\text{local}}=\frac{-F}{f}\bar{\Sigma}^{+}\left(x\right)\gamma^{\mu}\gamma_{5}\Sigma^{0}\left(x\right)\left(\partial_{\mu}+ie\mathscr{A}_{\mu}\left(x\right)\right)\pi^{+}\left(x\right).
\end{equation}
The corresponding nonlocal Lagrangian follows,
\begin{align}
	\mathcal{L}_{\Sigma\pi^+}^{\text{nl}} & =\frac{-F}{f} \int\dif^{4}y\,
	\bar{\Sigma}^{+}\left(x\right)\gamma^{\mu}\gamma_{5}\Sigma^{0}\left(x\right)\left(\partial_{\mu}+
	ie\int\dif^{4}a\mathscr{A}_{\mu}\left(x-a\right)F\left(a\right)\right)\nonumber\\
    & \times F\left(x-y\right)\exp\left[ie\int_{x}^{y}\dif
    z_{\nu}\int\dif^{4}b\mathscr{A}^{\nu}\left(z-b\right)
    F\left(b\right)\right]\pi^{+}\left(y\right),\label{eq:nonlocal}
\end{align}
where $F\left(x-y\right)$ is the correlation function. From the Lagrangian, one can see that the baryon fields are located at $x$, while the meson and photon fields are displaced. To make the Lagrangian gauge invariant locally, the gauge link $\exp\left[ie\int_{x}^{y}\dif z_{\nu}\int\dif^{4}b\mathscr{A}^{\nu}\left(z-b\right)F\left(b\right)\right]$ is introduced. 
Therefore, the photon could be emitted or annihilated from either the minimal substitution term, or the gauge link term.
Each photon field and meson field are associated with one correlation function.
With the correlation functions, the regulators and momentum dependence of the form factors at tree level,
could be generated automatically. In the numerical calculation, the correlation function in momentum space is chosen to have a dipole form.

The nonlocal baryon-photon interactions are obtained in the same procedure. 
For example, the local interaction between $\Sigma^{+}$ and photon is written as
\begin{equation}
	\mathcal{L}_{\text{EM}}^{\text{local}}=-e\bar{\Sigma}^{+}\left(x\right)\gamma^{\mu}\Sigma^{+}\left(x\right)\mathscr{A}_{\mu}\left(x\right)+\frac{\left(c_{1}+3c_{2}\right)e}{12m_{\Sigma}}\bar{\Sigma}^{+}\left(x\right)\sigma^{\mu\nu}\Sigma\left(x\right)^{+}\mathscr{F}_{\mu\nu}\left(x\right).
\end{equation}
The corresponding nonlocal Lagrangian is expressed as
\begin{equation}
	\mathcal{L}_{\text{EM}}^{\text{nl}}=-e\int\dif^{4}a\,
	\bar{\Sigma}^{+}\left(x\right)\gamma^{\mu}\Sigma\left(x\right)^{+}
	\mathscr{A}_{\mu}\left(x-a\right)F_{1}\left(a\right)+
	\frac{\left(c_{1}+3c_{2}\right)e}{6}\int\dif^{4}a\,
	\bar{\Sigma}^{+}\left(x\right)
	\frac{\sigma^{\mu\nu}}{2m_{\Sigma}}\Sigma\left(x\right)^{+}\mathscr{F}_{\mu\nu}\left(x-a\right)F_{2}\left(a\right),
\end{equation}
where $F_{1}\left(a\right)$ and $F_{2}\left(a\right)$ are the correlation functions for the nonlocal electric and magnetic interactions.

The momentum dependence of the form factors at tree level can be easily obtained with the Fourier transformation of the correlation function. As in our previous work \cite{He1,He2}, the correlation function is chosen such that the charge and magnetic form factors at tree level have the same momentum dependence as the baryon-meson vertex, i.e., $G_{M}^{B,\text{tree}}(q)=\mu_{B}G_{E}^{B,\text{tree}}(q)=\mu_{B}\tilde{F}(q)$,
where $\tilde{F}(q)$ is the Fourier transformation of the correlation function $F(a)$. Therefore, the corresponding functions $\tilde{F}_{1}\left(q\right)$ and $\tilde{F}_{2}\left(q\right)$ of $\Sigma^{+}$, for example,
are expressed as
\begin{equation}
	\tilde{F}_{1}^{\Sigma^{+}}\left(q\right)=\tilde{F}\left(q\right)\frac{12m_{\Sigma}^{2}+\left(3+c_{1}+3c_{2}\right)Q^{2}}{3\left(4m_{\Sigma}^{2}+Q^{2}\right)},\quad\tilde{F}_{2}^{\Sigma^{+}}\left(q\right)=\tilde{F}\left(q\right)\frac{4\left(c_{1}+3c_{2}\right)m_{\Sigma}^{2}}{3\left(4m_{\Sigma}^{2}+Q^{2}\right)},
\end{equation}
where $Q^{2}=-q^{2}$ is the momentum transfer. 

From Eq.\eqref{eq:nonlocal}, two kinds of $\Sigma$-$\pi$-photon vertices can be obtained. 
One is the normal coupling,
\begin{equation}
	\mathcal{L}^{\text{norm}}= -ie\frac{F}{f}\int\dif^{4}y  \bar{\Sigma}^{+}\left(x\right)\gamma^{\mu}\gamma_{5}\Sigma^{0}\left(x\right)\int\dif^{4}a\mathscr{A}_{\mu}\left(x-a\right)F\left(a\right)F\left(x-y\right)\pi^{+}\left(y\right).
\end{equation}
This interaction is similar to the traditional local Lagrangian, except the insertion of correlation functions.
The other is the additional interaction, obtained by the expansion of the gauge link,
\begin{equation}
	\mathcal{L}^{\text{add}}= -ie\frac{F}{f}\int\dif^{4}y  \bar{\Sigma}^{+}\left(x\right)\gamma^{\mu}\gamma_{5}\Sigma^{0}\left(x\right)\partial_{x,\mu}\left[F\left(x-y\right)\int_{x}^{y}\dif z_{\nu}\int\dif^{4}a\mathscr{A}^{\nu}\left(z-a\right)F\left(a\right)\pi^{+}\left(y\right)\right].
\end{equation}
The additional interaction is crucial to guarantee the charge conservation. 
With the nonlocal Lagrangian above, we can give a better description of the electromagnetic form factors.

\begin{figure}
	\includegraphics[scale=0.65]{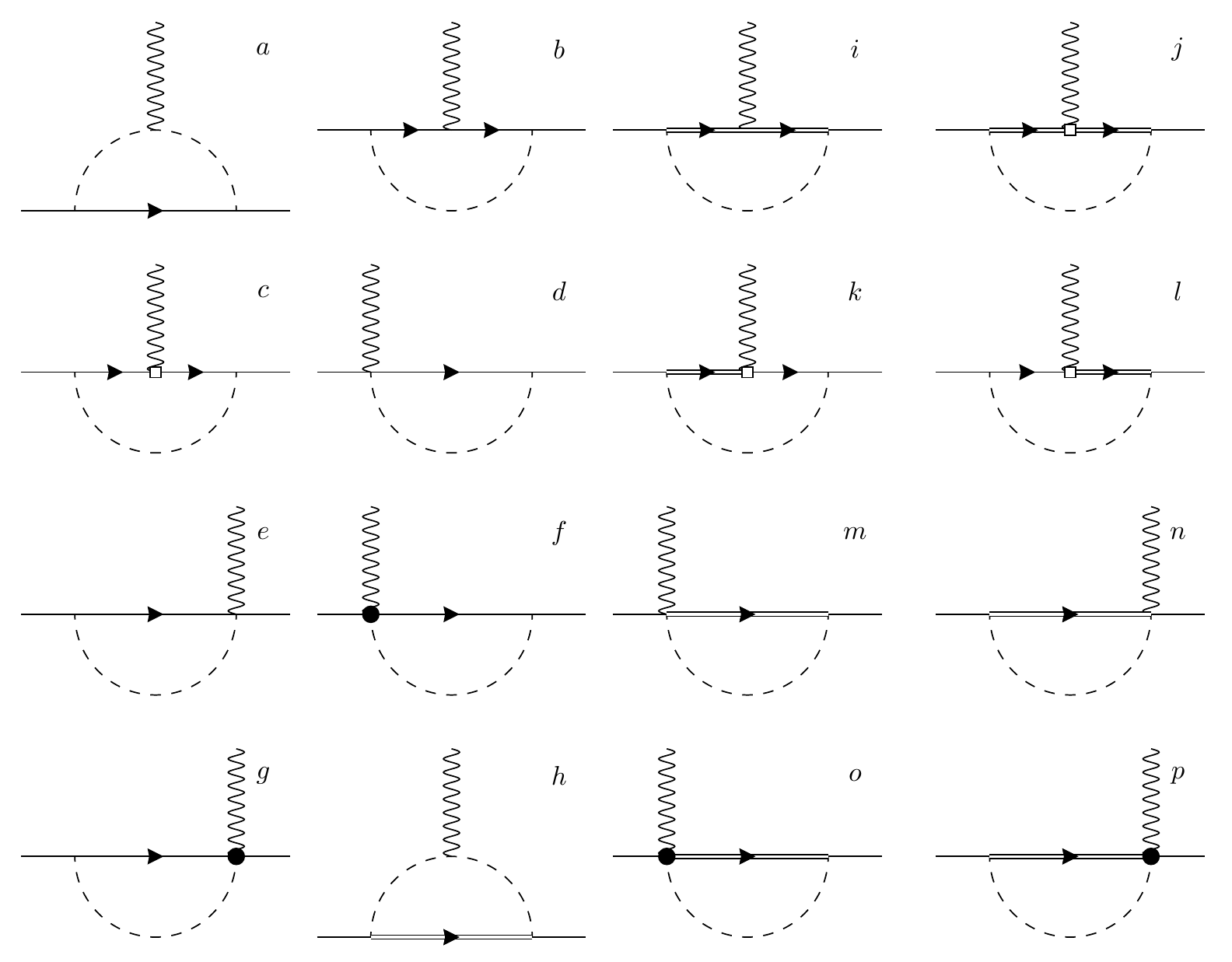}
	\caption{One-loop Feynman diagrams for the hyperon electromagnetic form factors.
		The solid, double-solid, dashed and wave lines are for the octet baryons,
		decuplet baryons, pseudo-scalar mesons and photons, respectively.
		The rectangle and black dot represent magnetic and additional interacting
		vertex.
		\label{fig:loop}
	}

\end{figure}
\begin{figure}
	\includegraphics[scale=0.6]{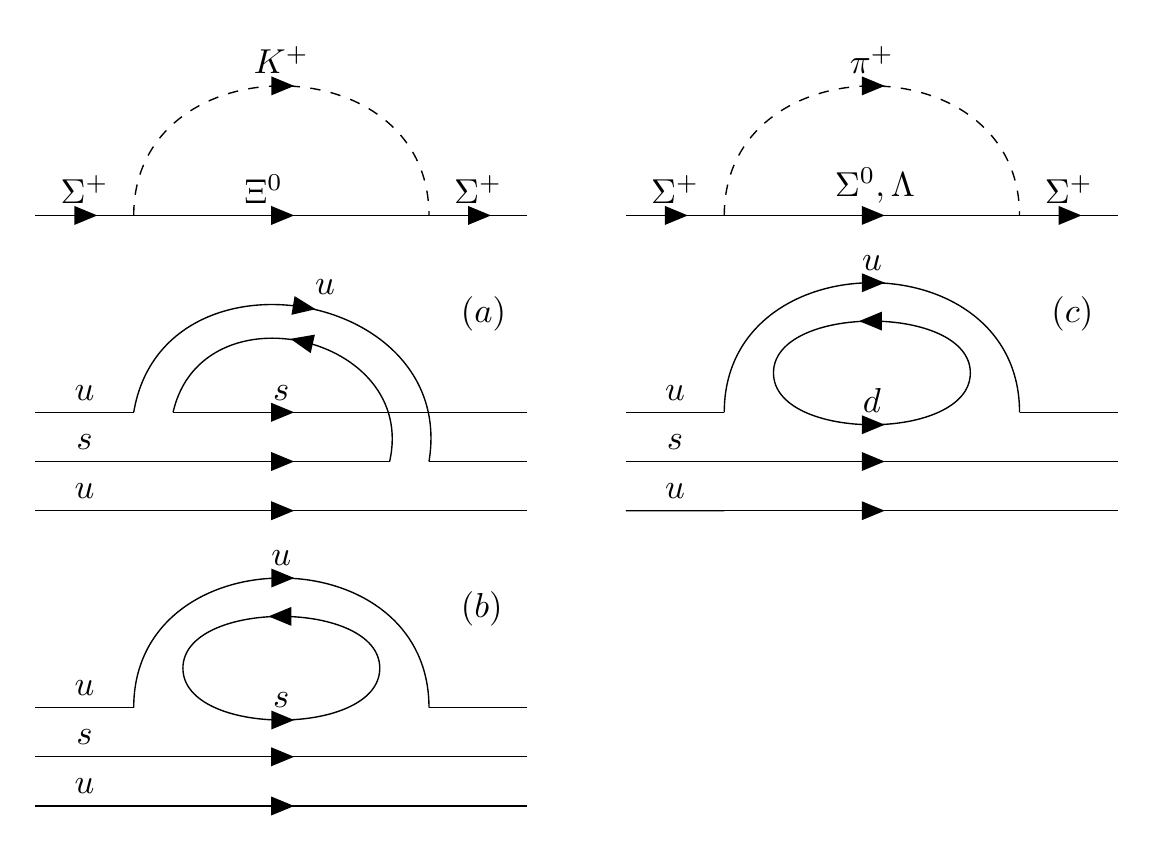}
	\caption{Quark flow diagrams for $K^{+}$ and $\pi^{+}$. (a) is the	quenched diagram for $K^{+}$. (b) and (c) are the disconnected sea diagrams for
		$K^{+}$ and $\pi^{+}$, respectively.
		\label{fig2}
	}
\end{figure}

\section{sea quark form factors}

The contributions of individual quark flavor $f$ $(f=u,d,s)$ to the Dirac and Pauli form factors of $\Sigma$ hyperons are defined as follow,
\begin{equation}
	\bra{\Sigma\left(p^{\prime}\right)}J_{\mu}^f\ket{\Sigma\left(p\right)}=\bar{u}\left(p^{\prime}\right)\left\{ \gamma_{\mu}F_{1}^{f,\Sigma}\left(Q^{2}\right)+\frac{i\sigma_{\mu\nu}q^{\nu}}{2m_{\Sigma}}F_{2}^{f,\Sigma}\left(Q^{2}\right)\right\} u\left(p\right),\label{eq:f1f2}
\end{equation}
where $q=p^{\prime}-p$. 
The electromagnetic form factors are defined as the combinations of the above form factors,
\begin{equation}
	G_{E}^{f,\Sigma}\left(Q^{2}\right)=F_{1}^{f,\Sigma}\left(Q^{2}\right)-\frac{Q^{2}}{4m_{\Sigma}^{2}}F_{2}^{f,\Sigma}\left(Q^{2}\right),\quad G_{M}^{f,\Sigma}\left(Q^{2}\right)=F_{1}^{f,\Sigma}\left(Q^{2}\right)+F_{2}^{f,\Sigma}\left(Q^{2}\right).
\end{equation}

According to the preceding Lagrangians, the one-loop Feynman diagrams which contribute to the $\Sigma$ electromagnetic form factors are shown in Fig.~\ref{fig:loop}. Further, we can evaluate the matrix element of Eq.~\eqref{eq:f1f2}.
The $\pi$ meson loops have the dominant contributions, while the contributions from $K$ meson loops are much smaller due to the large $K$ meson mass. The contributions from $\eta$ and $\eta'$ loops are even smaller,
which are neglected in our calculation. 
The inclusion of these mesons does not affect the main conclusion below.
In this section, we show the expressions of loop integrals for the
intermediate octet and decuplet baryons. As an example, we list the formulas for the contributions of $u$, $d$ and $s$ quark in $\Sigma^{+}$. For each diagram in Fig.~\ref{fig:loop}, there exist quenched and disconnected diagrams. In order to obtain the sea quark contribution, we need to find the coefficients for the disconnected diagrams. The coefficients for the quenched and disconnected loop diagrams can be extracted separately as in Ref.~\cite{Leinweber2004}, using the quark-flows of Fig.~\ref{fig2}. The obtained coefficients are the same as those extracted within the graded symmetry formalism in quenched chiral perturbation theory \cite{Bernard1992}.

In Fig.~\ref{fig2}, we plot the quark-flow diagrams for the rainbow diagram as an example, to show how to separate the quenched and sea quark contributions. The coefficients for the $K^+$ loop diagram in full QCD is $(D+F)^2/2f^2$. 
The coefficient of Fig.~\ref{fig2}b for the sea quark contribution is the same as that of Fig.~\ref{fig2}c for the $\pi+$ loop. Therefore, the coefficient for quenched sector can be obtained by subtracting the coefficient of sea diagram from the total one. The coefficients of $u$, $d$ and $s$ quark in $\Sigma$ for both quenched and sea quark-flow diagrams are listed in Table \ref{tab1}.

\begin{table}[tbp]
	\caption{\label{tab1}
		The coefficients of $u$, $d$ and $s$ quark in $\Sigma$,
		for both quenched and sea quark-flow diagrams.}	\centering
		\begin{tabular}{c|c|c|c|c}
			\hline
            Configurations    & $\Sigma-\pi$ & $\Lambda-\pi$ & $N-K$  & $\Xi-K$  
            \tabularnewline 
            \hline 
			$\Sigma_{u,\text{quench}}^{+}$ & $-\frac{D^{2}+9F^{2}}{6f^{2}}$ & $-\frac{D^{2}}{3f^{2}}$    &      &   $-\frac{(D+F)^{2}}{2f^{2}}$ \tabularnewline
			$\Sigma_{u,\text{sea}}^{+}$    & $\frac{D^{2}+3F^{2}}{6f^{2}}$  &   &   &  \tabularnewline
			 $\Sigma_{d,\text{sea}}^{+}$    & $\frac{F^{2}}{f^{2}}$    & $\frac{D^{2}}{3f^{2}}$   & $\frac{(D-F)^{2}}{2f^{2}}$    &   \tabularnewline
			\hline
			$\Sigma_{u,\text{quench}}^{0}$ & $-\frac{2D^{2}}{9f^{2}}$       & $-\frac{D^{2}}{6f^{2}}$         & $\frac{D^{2}-18DF+9F^{2}}{36f^{2}}$ & $-\frac{(D+F)^{2}}{4f^{2}}$ \tabularnewline
			$\Sigma_{u,\text{sea}}^{0}$    & $\frac{2D^{2}}{9f^{2}}$        & $\frac{D^{2}}{6f^{2}}$          & $\frac{2D^{2}}{9f^{2}}$  &   \tabularnewline	
			$\Sigma_{d,\text{quench}}^{0}$ & $-\frac{2D^{2}}{9f^{2}}$       & $-\frac{D^{2}}{6f^{2}}$         & $\frac{D^{2}-18DF+9F^{2}}{36f^{2}}$ & $-\frac{(D+F)^{2}}{4f^{2}}$ \tabularnewline
			$\Sigma_{d,\text{sea}}^{0}$ & $\frac{2D^{2}}{9f^{2}}$ & $\frac{D^{2}}{6f^{2}}$  & $\frac{2D^{2}}{9f^{2}}$ &  \tabularnewline  \hline
			$\Sigma_{s,\text{quench}}^{+}$ &  &  & $-\frac{(D-F)^{2}}{2f^{2}}$  & $\frac{D^{2}+6DF-3F^{2}}{6f^{2}}$  \tabularnewline
			$\Sigma_{s,\text{sea}}^{+}$  & &  &  & $\frac{D^{2}+3F^{2}}{3f^{2}}$  \tabularnewline
			$\Sigma_{s,\text{quench}}^{0}$  & & & $-\frac{(D-F)^{2}}{2f^{2}}$  & $\frac{D^{2}+18DF+9F^{2}}{18f^{2}}$  \tabularnewline
            $\Sigma_{s,\text{sea}}^{0}$ & & &    & $\frac{4D^{2}}{9f^{2}}$ \tabularnewline
             \hline
		\end{tabular}
\end{table}

With the obtained coefficients for the sea quark diagrams, the sea quark form factors of $u$, $d$ and $s$ in $\Sigma$ hyperons can be evaluated by summing up the contributions from each diagram in Fig.~\ref{fig:loop}. 
The contributions of Fig.~\ref{fig:loop}a are written as
\begin{align}
	\Gamma_{a,u}^{\mu}\left(\Sigma^{+}\right) & =\frac{D^{2}+3F^{2}}{6f^{2}}I_{a}^{\mu,\Sigma\pi},             \\
	\Gamma_{a,d}^{\mu}\left(\Sigma^{+}\right) & =\frac{D^{2}}{3f^{2}}I_{a}^{\mu,\Lambda\pi}+\frac{F^{2}}{f^{2}}I_{a}^{\mu,\Sigma\pi}+\frac{(D-F)^{2}}{2f^{2}}I_{a}^{\mu,NK}, \\
	\Gamma_{a,s}^{\mu}\left(\Sigma^{+}\right) & =\frac{D^{2}+3F^{2}}{3f^{2}}I_{a}^{\mu,\Xi K},
\end{align}
where the integral $I_{a}^{\mu,BM}$ is expressed as
\begin{equation}
	I_{a}^{\mu,BM}=\bar{u}\left(p^{\prime}\right)\tilde{F}\left(q\right)\int\frac{\dif^{4}k}{\left(2\pi\right)^{4}}\frac{\tilde{F}\left(q+k\right)\tilde{F}\left(k\right)}{D_{M}\left(k+q\right)}\frac{-\left(2k+q\right)^{\mu}}{D_{M}\left(k\right)}(\slashed{k}+\slashed{q})\gamma_{5}\frac{1}{\slashed{p}-\slashed{k}-m_{B}}\slashed{k}\gamma_{5}u\left(p\right).
\end{equation}
$D_{M}\left(k\right)$ is defined as
\[
	D_{M}\left(k\right)=k^{2}-m_{M}^{2}+i\varepsilon.
\]
$m_{B}$ and $m_{M}$ are the masses of the intermediate octet baryon $B$
and meson $M$, respectively.
The formulae for $\Sigma^0$ are similar, but with different coefficients.
The contributions of Fig.~\ref{fig:loop}b are expressed as
\begin{align}
	\Gamma_{b,u}^{\mu}\left(\Sigma^{+}\right) & =\frac{D^{2}\left(\left(c_{1}+3\left(c_{2}-c_{1}\right)+3\right)Q^{2}+12m_{\Sigma}^{2}\right)+6c_{1}DFQ^{2}+9F^{2}\left(\left(c_{2}+1\right)Q^{2}+4m_{\Sigma}^{2}\right)}{18f^{2}\left(4m_{\Sigma}^{2}+Q^{2}\right)}I_{b}^{\mu,\Sigma\pi},                                                \\
	\Gamma_{b,d}^{\mu}\left(\Sigma^{+}\right) & =\frac{D\left(\left(c_{1}+3\left(c_{2}-c_{1}\right)+3\right)DQ^{2}+3c_{1}FQ^{2}+12Dm_{\Lambda}^{2}\right)}{9f^{2}\left(4m_{\Lambda}^{2}+Q^{2}\right)}I_{b}^{\mu,\Lambda\pi}                                                                                                               \\
	                                          & +\frac{F\left(c_{1}DQ^{2}+3F\left(\left(c_{2}+1\right)Q^{2}+4m_{\Sigma}^{2}\right)\right)}{3f^{2}\left(4m_{\Sigma}^{2}+Q^{2}\right)}I_{b}^{\mu,\Sigma\pi}+\frac{(D-F)^{2}\left(\left(c_{2}-c_{1}\right)Q^{2}+4m_{N}^{2}+Q^{2}\right)}{2f^{2}\left(4m_{N}^{2}+Q^{2}\right)}I_{b}^{\mu,NK}, \\
	\Gamma_{b,s}^{\mu}\left(\Sigma^{+}\right) & =\frac{D^{2}\left(\left(c_{1}+3\left(c_{2}-c_{1}\right)+3\right)Q^{2}+12m_{\Xi}^{2}\right)+6c_{1}DFQ^{2}+9F^{2}\left(\left(c_{2}+1\right)Q^{2}+4m_{\Xi}^{2}\right)}{9f^{2}\left(4m_{\Xi}^{2}+Q^{2}\right)}I_{b}^{\mu,\Xi K},
\end{align}
where the integral $I_{b}^{\mu,BM}$ is written as
\begin{equation}
	I_{b}^{\mu,BM}=\bar{u}\left(p^{\prime}\right)\tilde{F}\left(q\right)\int\frac{\dif^{4}k}{(2\pi)^{4}}\frac{\tilde{F}\left(k\right)^{2}}{D_{M}\left(k\right)}\slashed{k}\gamma_{5}\frac{1}{\slashed{p}^{\prime}-\slashed{k}-m_{B}}\gamma^{\mu}\frac{1}{\slashed{p}-\slashed{k}-m_{B}}\slashed{k}\gamma_{5}u\left(p\right).
\end{equation}
Fig.~\ref{fig:loop}c is similar to Fig.~\ref{fig:loop}b, except the former is for the magnetic interaction. 
The contributions of this diagram are written
as
\begin{align}
	\Gamma_{c,u}^{\mu}\left(\Sigma^{+}\right) & =\frac{im_{\Sigma}\left(3\left(c_{2}-c_{1}\right)\left(D^{2}+3F^{2}\right)+c_{1}(D+3F)^{2}\right)}{9f^{2}\left(4m_{\Sigma}^{2}+Q^{2}\right)}I_{c}^{\mu,\Sigma\pi},                                                                                                                \\
	\Gamma_{c,d}^{\mu}\left(\Sigma^{+}\right) & =\frac{2iDm_{\Lambda}\left(c_{1}(D+3F)+3\left(c_{2}-c_{1}\right)D\right)}{9f^{2}\left(4m_{\Lambda}^{2}+Q^{2}\right)}I_{c}^{\mu,\Lambda\pi}+\frac{2iFm_{\Sigma}\left(c_{1}(D+3F)+3\left(c_{2}-c_{1}\right)F\right)}{3f^{2}\left(4m_{\Sigma}^{2}+Q^{2}\right)}I_{c}^{\mu,\Sigma\pi} \\
	                                          & +\frac{i\left(c_{2}-c_{1}\right)(D-F)^{2}m_{N}}{f^{2}\left(4m_{N}^{2}+Q^{2}\right)}I_{c}^{\mu,NK},                                   \\
	\Gamma_{c,s}^{\mu}\left(\Sigma^{+}\right) & =\frac{2im_{\Xi}\left(3\left(c_{2}-c_{1}\right)\left(D^{2}+3F^{2}\right)+c_{1}(D+3F)^{2}\right)}{9f^{2}\left(4m_{\Xi}^{2}+Q^{2}\right)}I_{c}^{\mu,\Xi K},
\end{align}
where $I_{c}^{\mu,BM}$ is,
\begin{equation}
	I_{c}^{\mu,BM}=\bar{u}\left(p^{\prime}\right)\tilde{F}\left(q\right)\int\frac{\dif^{4}k}{\left(2\pi\right)^{4}}\frac{\tilde{F}\left(k\right)^{2}}{D_{M}\left(k\right)}\slashed{k}\gamma_{5}\frac{1}{\slashed{p}^{\prime}-\slashed{k}-m_{B}}i\sigma^{\mu\nu}q_{\nu}\frac{1}{\slashed{p}-\slashed{k}-m_{B}}\slashed{k}\gamma_{5}u\left(p\right).
\end{equation}
Figs.~\ref{fig:loop}d and \ref{fig:loop}e are the Kroll-Ruderman
diagrams. The contributions of these two diagrams are written as
\begin{align}
	\Gamma_{d+e,u}^{\mu}\left(\Sigma^{+}\right) & =\frac{D^{2}+3F^{2}}{6f^{2}}I_{d+e}^{\mu,\Sigma\pi},                                     \\
	\Gamma_{d+e,d}^{\mu}\left(\Sigma^{+}\right) & =\frac{D^{2}}{3f^{2}}I_{d+e}^{\mu,\Lambda\pi}+\frac{F^{2}}{f^{2}}I_{d+e}^{\mu,\Sigma\pi}+\frac{(D-F)^{2}}{2f^{2}}I_{d+e}^{\mu,NK}, \\
	\Gamma_{d+e,s}^{\mu}\left(\Sigma^{+}\right) & =\frac{D^{2}+3F^{2}}{3f^{2}}I_{d+e}^{\mu,\Xi K},
\end{align}
where
\begin{equation}
	I_{d+e}^{\mu,BM}=\bar{u}\left(p^{\prime}\right)\tilde{F}\left(q\right)\int\frac{\dif^{4}k}{\left(2\pi\right)^{4}}\frac{\tilde{F}\left(k\right)^{2}}{D_{M}\left(k\right)}\Big\lbrace\slashed{k}\gamma_{5}\frac{1}{\slashed{p}^{\prime}-\slashed{k}-m_{B}}\gamma^{\mu}\gamma_{5}+\gamma^{\mu}\gamma_{5}\frac{1}{\slashed{p}-\slashed{k}-m_{B}}\slashed{k}\gamma_{5}\Big\rbrace u\left(p\right).
\end{equation}
Figs.~\ref{fig:loop}f and \ref{fig:loop}g are the additional diagrams
which are generated from the expansion of the gauge link. The
contributions of these two diagrams are expressed as
\begin{align}
	\Gamma_{f+g,u}^{\mu}\left(\Sigma^{+}\right) & =\frac{D^{2}+3F^{2}}{6f^{2}}I_{f+g}^{\mu,\Sigma\pi},                                           \\
	\Gamma_{f+g,d}^{\mu}\left(\Sigma^{+}\right) & =\frac{D^{2}}{3f^{2}}I_{f+g}^{\mu,\Lambda\pi}+\frac{F^{2}}{f^{2}}I_{f+g}^{\mu,\Sigma\pi}+\frac{(D-F)^{2}}{2f^{2}}I_{f+g}^{\mu,NK}, \\
	\Gamma_{f+g,s}^{\mu}\left(\Sigma^{+}\right) & =\frac{D^{2}+3F^{2}}{3f^{2}}I_{f+g}^{\mu,\Xi K},
\end{align}
where
\begin{align}
	I_{f+g}^{\mu,BM} & =\bar{u}\left(p^{\prime}\right)\tilde{F}\left(q\right)\int\frac{\dif^{4}k}{\left(2\pi\right)^{4}}\frac{\tilde{F}\left(k\right)}{D_{M}\left(k\right)}\Big\lbrace\frac{\left(2k-q\right)^{\mu}}{2kq-q^{2}}\left[\tilde{F}\left(k-q\right)-\tilde{F}\left(k\right)\right]\slashed{k}\gamma_{5}\frac{1}{\slashed{p}^{\prime}-\slashed{k}-m_{B}}\left(-\slashed{k}+\slashed{q}\right)\gamma_{5}\nonumber \\
	                 & +\frac{\left(2k+q\right)^{\mu}}{2kq+q^{2}}\left[\tilde{F}\left(k+q\right)-\tilde{F}\left(k\right)\right]\left(\slashed{k}+\slashed{q}\right)\gamma_{5}\frac{1}{\slashed{p}-\slashed{k}-m_{B}}\slashed{k}\gamma_{5}\Big\rbrace u\left(p\right).
\end{align}

Now we show the expressions of one loop integrals for decuplet intermediate
states. The contributions for Fig.~\ref{fig:loop}h can be written as
\begin{align}
	\Gamma_{h,u}^{\mu}\left(\Sigma^{+}\right) & =\frac{\mathcal{C}^{2}}{24f^{2}}I_{h}^{\mu,\Sigma^{\ast}\pi}+\frac{\mathcal{C}^{2}}{6f^{2}}I_{h}^{\mu,\Delta K}, \\
	\Gamma_{h,d}^{\mu}\left(\Sigma^{+}\right) & =\frac{\mathcal{C}^{2}}{12f^{2}}I_{h}^{\mu,\Sigma^{\ast}\pi}+\frac{\mathcal{C}^{2}}{6f^{2}}I_{h}^{\mu,\Delta K}, \\
	\Gamma_{h,s}^{\mu}\left(\Sigma^{+}\right) & =\frac{\mathcal{C}^{2}}{12f^{2}}I_{h}^{\mu,\Xi^{\ast}K},
\end{align}
where the integral $I_{h}^{\mu,TM}$ is expressed as
\begin{align}
	I_{h}^{\mu,TM} & =\bar{u}\left(p^{\prime}\right)\tilde{F}\left(q\right)\int\frac{\dif^{4}k}{\left(2\pi\right)^{4}}\frac{\tilde{F}\left(q+k\right)\tilde{F}\left(k\right)}{D_{M}\left(k\right)}\frac{2\left(k+q\right)^{\mu}}{D_{M}\left(q+k\right)}\nonumber \\
	               & \times\left(\left(k+q\right)^{\sigma}+z\left(\slashed{k}+\slashed{q}\right)\gamma^{\sigma}\right)\frac{1}{\slashed{p}-\slashed{k}-m_{T}}S_{\sigma\rho}\left(p-k\right)\left(-k^{\rho}-z\gamma^{\rho}\slashed{k}\right)u\left(p\right).
\end{align}
$m_{T}$ is the mass of the decuplet intermediate state and $S_{\sigma\rho}\left(p\right)$
is expressed as
\[
	S_{\sigma\rho}\left(p\right)=-g_{\sigma\rho}+\frac{\gamma_{\sigma}\gamma_{\rho}}{3}+\frac{p_{\sigma}p_{\rho}}{3m_{T}^{2}}+\frac{\gamma_{\sigma}p_{\rho}-\gamma_{\rho}p_{\sigma}}{3m_{T}}.
\]
The contributions for Fig.~\ref{fig:loop}i are written as
\begin{align}
	\Gamma_{i,u}^{\mu}\left(\Sigma^{+}\right) & =\frac{\mathcal{C}^{2}\left(\left(c_{1}+3c_{2}+3\right)Q^{2}+12m_{\Sigma^{*}}^{2}\right)}{72f^{2}\left(4m_{\Sigma^{*}}^{2}+Q^{2}\right)}I_{i}^{\mu,\Sigma^{\ast}\pi}+\frac{\mathcal{C}^{2}\left(\left(c_{1}+3c_{2}+3\right)Q^{2}+12m_{\Delta}^{2}\right)}{18f^{2}\left(4m_{\Delta}^{2}+Q^{2}\right)}I_{i}^{\mu,\Delta K}, \\
	\Gamma_{i,d}^{\mu}\left(\Sigma^{+}\right) & =\frac{\mathcal{C}^{2}\left(\left(c_{1}+3c_{2}+3\right)Q^{2}+12m_{\Sigma^{*}}^{2}\right)}{36f^{2}\left(4m_{\Sigma^{*}}^{2}+Q^{2}\right)}I_{i}^{\mu,\Sigma^{\ast}\pi}+\frac{\mathcal{C}^{2}\left(\left(c_{1}+3c_{2}+3\right)Q^{2}+12m_{\Delta}^{2}\right)}{18f^{2}\left(4m_{\Delta}^{2}+Q^{2}\right)}I_{i}^{\mu,\Delta K}, \\
	\Gamma_{i,s}^{\mu}\left(\Sigma^{+}\right) & =\frac{\mathcal{C}^{2}\left(\left(c_{1}+3c_{2}+3\right)Q^{2}+12m_{\Xi^{*}}^{2}\right)}{36f^{2}\left(4m_{\Xi^{*}}^{2}+Q^{2}\right)}I_{i}^{\mu,\Xi^{\ast}K},
\end{align}
where the integral $I_{i}^{\mu,TM}$ is written as
\begin{align}
	I_{i}^{\mu,TM} & =\bar{u}\left(p^{\prime}\right)\tilde{F}\left(q\right)\int\frac{\dif^{4}k}{\left(2\pi\right)^{4}}\frac{\tilde{F}\left(k\right)^{2}}{D_{M}\left(k\right)}\left(k^{\sigma}+z\slashed{k}\gamma^{\sigma}\right)\nonumber                                      \\
	               & \times\frac{1}{\slashed{p}^{\prime}-\slashed{k}-m_{T}}S_{\sigma\alpha}\left(p^{\prime}-k\right)\gamma^{\alpha\beta\mu}\frac{1}{\slashed{p}-\slashed{k}-m_{T}}S_{\beta\rho}\left(p-k\right)\left(k^{\rho}+z\gamma^{\rho}\slashed{k}\right)u\left(p\right).
\end{align}
The contributions for Fig.~\ref{fig:loop}j are written as
\begin{align}
	\Gamma_{j,u}^{\mu}\left(\Sigma^{+}\right) & =-\frac{i\left(c_{1}+3c_{2}\right)\mathcal{C}^{2}m_{\Sigma^{*}}}{36f^{2}\left(4m_{\Sigma^{*}}^{2}+Q^{2}\right)}I_{j}^{\mu,\Sigma^{\ast}\pi}-\frac{i\left(c_{1}+3c_{2}\right)\mathcal{C}^{2}m_{\Delta}}{9f^{2}\left(4m_{\Delta}^{2}+Q^{2}\right)}I_{j}^{\mu,\Delta K}, \\
	\Gamma_{j,d}^{\mu}\left(\Sigma^{+}\right) & =-\frac{i\left(c_{1}+3c_{2}\right)\mathcal{C}^{2}m_{\Sigma^{*}}}{18f^{2}\left(4m_{\Sigma^{*}}^{2}+Q^{2}\right)}I_{j}^{\mu,\Sigma^{\ast}\pi}-\frac{i\left(c_{1}+3c_{2}\right)\mathcal{C}^{2}m_{\Delta}}{9f^{2}\left(4m_{\Delta}^{2}+Q^{2}\right)}I_{j}^{\mu,\Delta K}, \\
	\Gamma_{j,s}^{\mu}\left(\Sigma^{+}\right) & =-\frac{i\left(c_{1}+3c_{2}\right)\mathcal{C}^{2}m_{\Xi^{*}}}{18f^{2}\left(4m_{\Xi^{*}}^{2}+Q^{2}\right)}I_{j}^{\mu,\Xi^{\ast}K},
\end{align}
where the integral $I_{j}^{\mu,TM}$ is expressed as
\begin{align}
	I_{j}^{\mu,TM} & =\bar{u}\left(p^{\prime}\right)\tilde{F}\left(q\right)\int\frac{\dif^{4}k}{\left(2\pi\right)^{4}}\frac{\tilde{F}\left(k\right)^{2}}{D_{M}\left(k\right)}\left(k^{\sigma}+z\slashed{k}\gamma^{\sigma}\right)\nonumber                                         \\
	               & \times\frac{1}{\slashed{p}^{\prime}-\slashed{k}-m_{T}}S_{\sigma\nu}\left(p^{\prime}-k\right)i\sigma^{\mu\lambda}q_{\lambda}\frac{1}{\slashed{p}-\slashed{k}-m_{T}}S^{\nu\rho}\left(p-k\right)\left(k_{\rho}+z\gamma_{\rho}\slashed{k}\right)u\left(p\right).
\end{align}
The contributions for the intermediate octet-decuplet transition diagrams
Figs.~\ref{fig:loop}k and \ref{fig:loop}l are expressed as
\begin{align}
	\Gamma_{k+l,u}^{\mu}\left(\Sigma^{+}\right) & =\frac{c_{1}\mathcal{C}(D-F)}{24f^{2}m_{\Sigma}}I_{k+l}^{\mu,\Sigma^{\ast}\Sigma\pi},                                                  \\
	\Gamma_{k+l,d}^{\mu}\left(\Sigma^{+}\right) & =\frac{c_{1}\mathcal{C}D}{12f^{2}m_{\Lambda}}I_{k+l}^{\mu,\Sigma^{\ast}\Lambda\pi}-\frac{c_{1}\mathcal{C}F}{12f^{2}m_{\Sigma}}I_{k+l}^{\mu,\Sigma^{\ast}\Sigma\pi}+\frac{c_{1}\mathcal{C}(D-F)}{6f^{2}m_{N}}I_{k+l}^{\mu,\Delta N\pi}, \\
	\Gamma_{k+l,s}^{\mu}\left(\Sigma^{+}\right) & =\frac{c_{1}\mathcal{C}(D-F)}{12f^{2}m_{\Xi}}I_{k+l}^{\mu,\Xi^{*}\Xi K},
\end{align}
where the integral $I_{k+l}^{\mu,TBM}$ is written as
\begin{align}
	I_{k+l}^{\mu,TBM} & =\bar{u}\left(p^{\prime}\right)\tilde{F}\left(q\right)\int\frac{\dif^{4}k}{\left(2\pi\right)^{4}}\frac{\tilde{F}\left(k\right)^{2}}{D_{M}\left(k\right)}\Big\lbrace\slashed{k}\gamma_{5}\frac{1}{\slashed{p}^{\prime}-\slashed{k}-m_{B}}\left(-\slashed{q}\gamma_{5}\right)\frac{1}{\slashed{p}-\slashed{k}-m_{T}}S^{\mu\rho}\left(p-k\right)\left(k_{\rho}+z\gamma_{\rho}\slashed{k}\right)\nonumber \\
	                  & +\slashed{k}\gamma_{5}\frac{1}{\slashed{p}^{\prime}-\slashed{k}-m_{B}}\gamma^{\mu}\gamma_{5}q_{\nu}\frac{1}{\slashed{p}-\slashed{k}-m_{T}}S^{\nu\rho}\left(p-k\right)\left(k_{\rho}+z\gamma_{\rho}\slashed{k}\right)\nonumber                                        \\
	                  & +\left(k_{\nu}+z\slashed{k}\gamma_{\nu}\right)\frac{1}{\slashed{p}^{\prime}-\slashed{k}-m_{T}}S^{\nu\rho}\left(p^{\prime}-k\right)\left(-q_{\rho}\gamma^{\mu}\gamma_{5}\right)\frac{1}{\slashed{p}-\slashed{k}-m_{B}}\slashed{k}\gamma_{5}\nonumber                                        \\
	                  & +\left(k_{\nu}+z\slashed{k}\gamma_{\nu}\right)\frac{1}{\slashed{p}^{\prime}-\slashed{k}-m_{T}}S^{\nu\mu}\left(p^{\prime}-k\right)\slashed{q}\gamma_{5}\frac{1}{\slashed{p}-\slashed{k}-m_{B}}\slashed{k}\gamma_{5}\Big\rbrace u\left(p\right).
\end{align}
The contributions for the Kroll-Ruderman diagrams Figs.~\ref{fig:loop}m
and \ref{fig:loop}n are written as
\begin{align}
	\Gamma_{m+n,u}^{\mu}\left(\Sigma^{+}\right) & =\frac{\mathcal{C}^{2}}{24f^{2}}I_{m+n}^{\mu,\Sigma^{\ast}\pi}+\frac{\mathcal{C}^{2}}{6f^{2}}I_{m+n}^{\mu,\Delta K}, \\
	\Gamma_{m+n,d}^{\mu}\left(\Sigma^{+}\right) & =\frac{\mathcal{C}^{2}}{12f^{2}}I_{m+n}^{\mu,\Sigma^{\ast}\pi}+\frac{\mathcal{C}^{2}}{6f^{2}}I_{m+n}^{\mu,\Delta K}, \\
	\Gamma_{m+n,s}^{\mu}\left(\Sigma^{+}\right) & =\frac{\mathcal{C}^{2}}{12f^{2}}I_{m+n}^{\mu,\Xi^{\ast}K},
\end{align}
where the integral $I_{m+n}^{\mu,TM}$ is written as
\begin{align}
	 & I_{m+n}^{\mu,TM}=\bar{u}\left(p^{\prime}\right)\tilde{F}\left(q\right)\int\frac{\dif^{4}k}{\left(2\pi\right)^{4}}\frac{\tilde{F}\left(k\right)^{2}}{D_{M}\left(k\right)}\Big\lbrace\left(k_{\sigma}+z\slashed{k}\gamma_{\sigma}\right)\frac{1}{\slashed{p}^{\prime}-\slashed{k}-m_{T}}S^{\sigma\rho}\left(p^{\prime}-k\right)\left(g_{\rho}^{\mu}+z\gamma_{\rho}\gamma^{\mu}\right)+ \\
	 & \left(g_{\sigma}^{\mu}+z\gamma^{\mu}\gamma_{\sigma}\right)\frac{1}{\slashed{p}-\slashed{k}-m_{T}}S^{\sigma\rho}\left(p-k\right)\left(k_{\rho}+z\gamma_{\rho}\slashed{k}\right)\Big\rbrace u\left(p\right).
\end{align}
Finally, the contributions for the additional diagrams with intermediate decuplet
states Figs.~\ref{fig:loop}o and \ref{fig:loop}p are expressed as
\begin{align}
	\Gamma_{o+p,u}^{\mu}\left(\Sigma^{+}\right) & =\frac{\mathcal{C}^{2}}{24f^{2}}I_{o+p}^{\mu,\Sigma^{\ast}\pi}+\frac{\mathcal{C}^{2}}{6f^{2}}I_{o+p}^{\mu,\Delta K}, \\
	\Gamma_{o+p,d}^{\mu}\left(\Sigma^{+}\right) & =  \frac{\mathcal{C}^{2}}{12f^{2}}I_{o+p}^{\mu,\Sigma^{\ast}\pi}+\frac{\mathcal{C}^{2}}{6f^{2}}I_{o+p}^{\mu,\Delta K},   \\
	\Gamma_{o+p,s}^{\mu}\left(\Sigma^{+}\right) & =\frac{\mathcal{C}^{2}}{12f^{2}}I_{o+p}^{\mu,\Xi^{\ast}K},
\end{align}
where the integral $I_{o+p}^{\mu,TM}$ is written as
\begin{align}
	I_{o+p}^{\mu,TM} & =\bar{u}\left(p^{\prime}\right)\tilde{F}\left(q\right)\int\frac{\dif^{4}k}{\left(2\pi\right)^{4}}\frac{\tilde{F}\left(k\right)}{D_{M}\left(k\right)}\nonumber       \\
	                 & \Big\lbrace\frac{\left(-2k+q\right)^{\mu}}{-2kq+q^{2}}\left(\tilde{F}\left(k-q\right)-\tilde{F}\left(k\right)\right)\left(k_{\sigma}+z\slashed{k}\gamma_{\sigma}\right)\frac{1}{\slashed{p}^{\prime}-\slashed{k}-m_{T}}S^{\sigma\rho}\left(p^{\prime}-k\right)\left(\left(k-q\right)_{\rho}+z\gamma_{\rho}\left(\slashed{k}-\slashed{q}\right)\right)\nonumber \\
	                 & +\frac{\left(2k+q\right)^{\mu}}{2kq+q^{2}}\left(\tilde{F}\left(k+q\right)-\tilde{F}\left(k\right)\right)\left(\left(k+q\right)_{\sigma}+z\left(\slashed{k}+\slashed{q}\right)\gamma_{\sigma}\right)\frac{1}{\slashed{p}-\slashed{k}-m_{T}}S^{\sigma\rho}\left(p-k\right)\left(k_{\rho}+z\gamma_{\rho}\slashed{k}\right)\Big\rbrace u(p).
\end{align}
Using Package-X \cite{tool.packagex} to simplify the loop integral, we have gathered the results for the electromagnetic form factors.

\begin{figure}[H]
	\begin{center}
		\includegraphics[scale=0.6]{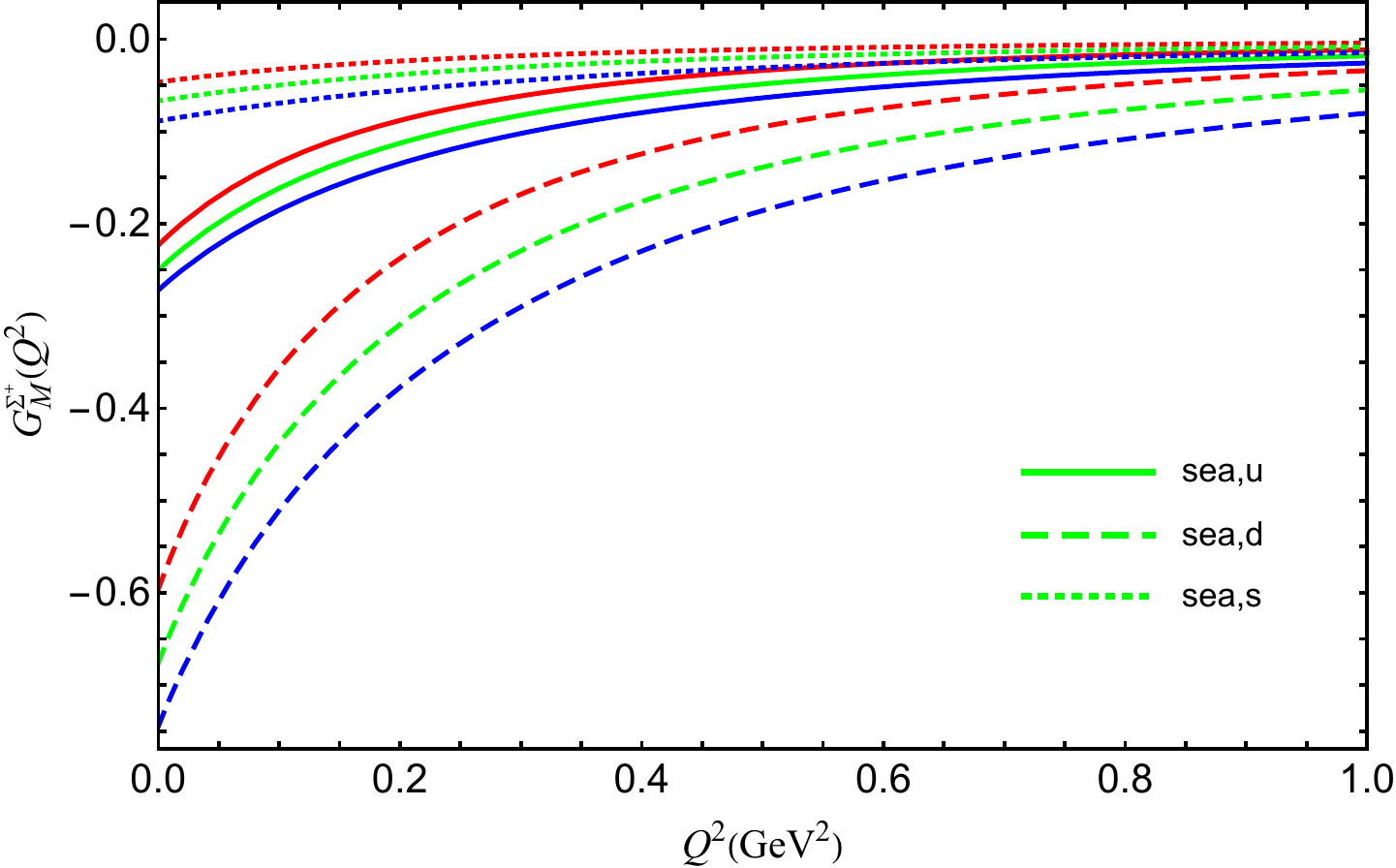}
		\caption{\label{fig:Sigma+gm}
			The sea quark contributions to the magnetic form factors of $\Sigma^{+}$ hyperon versus momentum transfer $Q^{2}$. The solid, dashed and dotted lines are for the contributions from $u$--sea, $d$--sea and $s$--sea, respectively. The lines with color red, green and blue are for $\Lambda=0.8$, 0.9 and 1.0 GeV, respectively.}
	\end{center}
\end{figure}

\section{numerical results}

The coupling constants between octet baryons and mesons are determined by two parameters $D$ and $F$.
In the numerical calculations, the parameters are chosen to be $D=0.76$ and $F=0.50$ ($g_{A}=D+F=1.26$)
\cite{Borasoy}. The coupling constant $\mathcal{C}$ is chosen to
be $1.0$ which is the same as in Refs.~\cite{He1,He2}. The off-shell
parameter $z$ is $-1$ \cite{nath.decuplet}. 
Besides, the physical masses are employed, for mesons, octet and decuplet baryons. 
The covariant regulator is chosen to be a dipole form \cite{He2,Yang}
\begin{equation}
	\tilde{F}\left(k\right)=\frac{\Lambda^{4}}{(k^{2}-m_{j}^{2}-\Lambda^{2})^{2}},
\end{equation}
where $m_{j}$ is the meson mass for the baryon-meson interaction
and it is zero for the hadron-photon interaction. It was found that
when $\Lambda$ around 0.9 GeV, the obtained electromagnetic and strange form
factors were very close to the experimental data and lattice simulation. 
The low energy constants $c_{1}$ and $c_{2}$ are determined by fitting
the experimental magnetic moments of octet baryons. 
They are found to be $1.736$ and $0.329$, 
which give the minimal $\chi^{2}$ of the octet magnetic moments.

\begin{figure}[H]
	\begin{center}
		\includegraphics[scale=0.6]{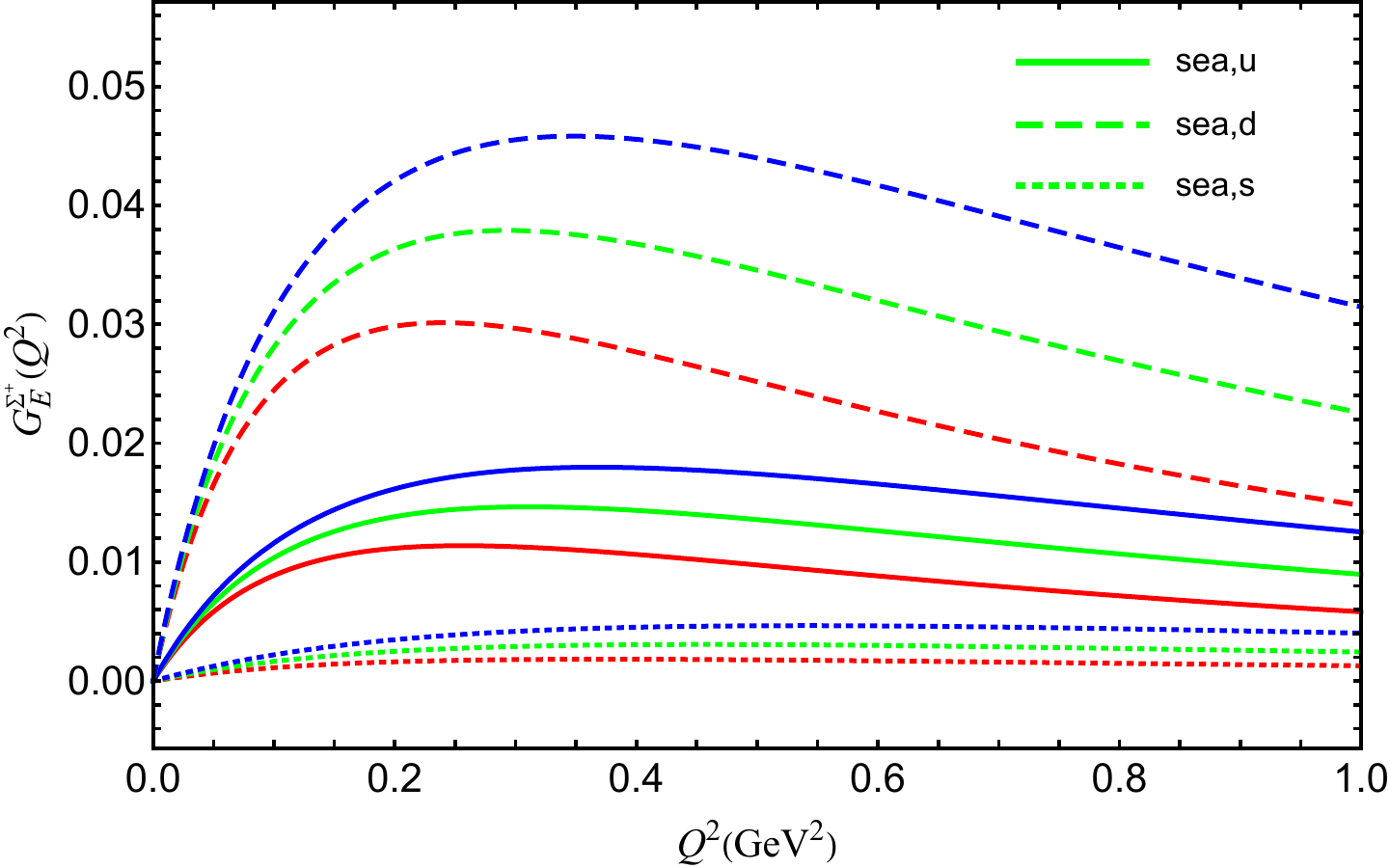}
		\caption{\label{fig:Sigma+ge}
			Same as Fig.~\ref{fig:Sigma+gm} but for the electric form factors.}
	\end{center}
\end{figure}

The sea quark contributions to the magnetic form factor of $\Sigma^{+}$ baryon versus momentum
transfer $Q^{2}$ are plotted in Fig.~\ref{fig:Sigma+gm}. 
The solid, dashed and dotted lines are for $u$, $d$ and $s$ quark with unit charge, respectively. 
The lines colored red, green and blue are for $\Lambda=0.8$, $0.9$ and $1.0$ \si{GeV}, respectively. 
From the figure, one can see all the sea quark contributions to the magnetic form factors of $\Sigma^+$ are negative.
When $Q^2=0$, the sea quark magnetic moments $\mu^{u,\Sigma^+}$, $\mu^{d,\Sigma^+}$
and $\mu^{s,\Sigma^+}$ are $-0.251 \pm 0.025$, $-0.676 \pm 0.083$ and $-0.067 \pm 0.022$, respectively. 
They remain negative, and their magnitudes decrease with the increasing momentum transfer. 
Compared with the sea quark contribution in the nucleon,
$G_M^{f,\Sigma^+}$ is larger than the corresponding $G_M^{f,N}$.
It is expected the light quark contribution is much larger than the strange quark contribution, 
as in the nucleon case. The magnitude of $G_M^{d,\Sigma^+}$ is the largest one, 
and it is much larger than $G_M^{u,\Sigma^+}$. 
Similarly as in the nucleon, the asymmetry of $u$ sea and $d$ sea is obvious. 
This flavor asymmetry is not raised by the masses difference of $u$ and $d$ quark. 
It is result from the different numbers of valence quark $u$ and $d$ in $\Sigma^+$. 
For example, in Fig.~\ref{fig:loop} there is only $\pi^+$ loop 
which contributes to the form factor of $d$ sea quark. 
No $\pi^-$ diagram exists which exactly contributes to the form factor of $u$ sea quark. 
It is interesting to see, whether this sea quark asymmetry could be obtained from lattice simulation. 
Currently, lattice simulations have not shown this asymmetry in proton, 
where the light quark form factors are somewhat like the average of the $u$ and $d$ contributions. 
Since the difference between $u$ sea and $d$ sea is very large, 
to make the direct comparison between EFT calculation and lattice simulation meaningful, 
we will show the light sea quark contributions to the form factors of $\Sigma^0$ hyperon latter, 
where the sea contributions from $u$ and $d$ quark are the same.

In $\Sigma^+$ hyperon, there is no valence $d$ quark. All the contribution of $d$ quark is from the sea. 
This is comparable with the strange quark contribution in nucleon. 
The difference is that, $G_M^{d,\Sigma^+}$ is generated from the $\pi$ loop, 
while $G_M^{s,N}$ is from the $K$ loop. 
As a result, $G_M^{d,\Sigma^+}$ is much larger than the strange magnetic form factor of nucleon,
which make it easier for lattice simulation to acquire an unambiguous sign, 
for this pure sea quark contribution. 
Once lattice confirms that $G_M^{d,\Sigma^+}$ is negative, 
it should suggest the meson-loop scenario, 
i.e. the anti-quark in the nucleon should be in the state with orbital angular momentum 1 \cite{Zou}.

\begin{figure}[H]
	\begin{center}
		\includegraphics[scale=0.6]{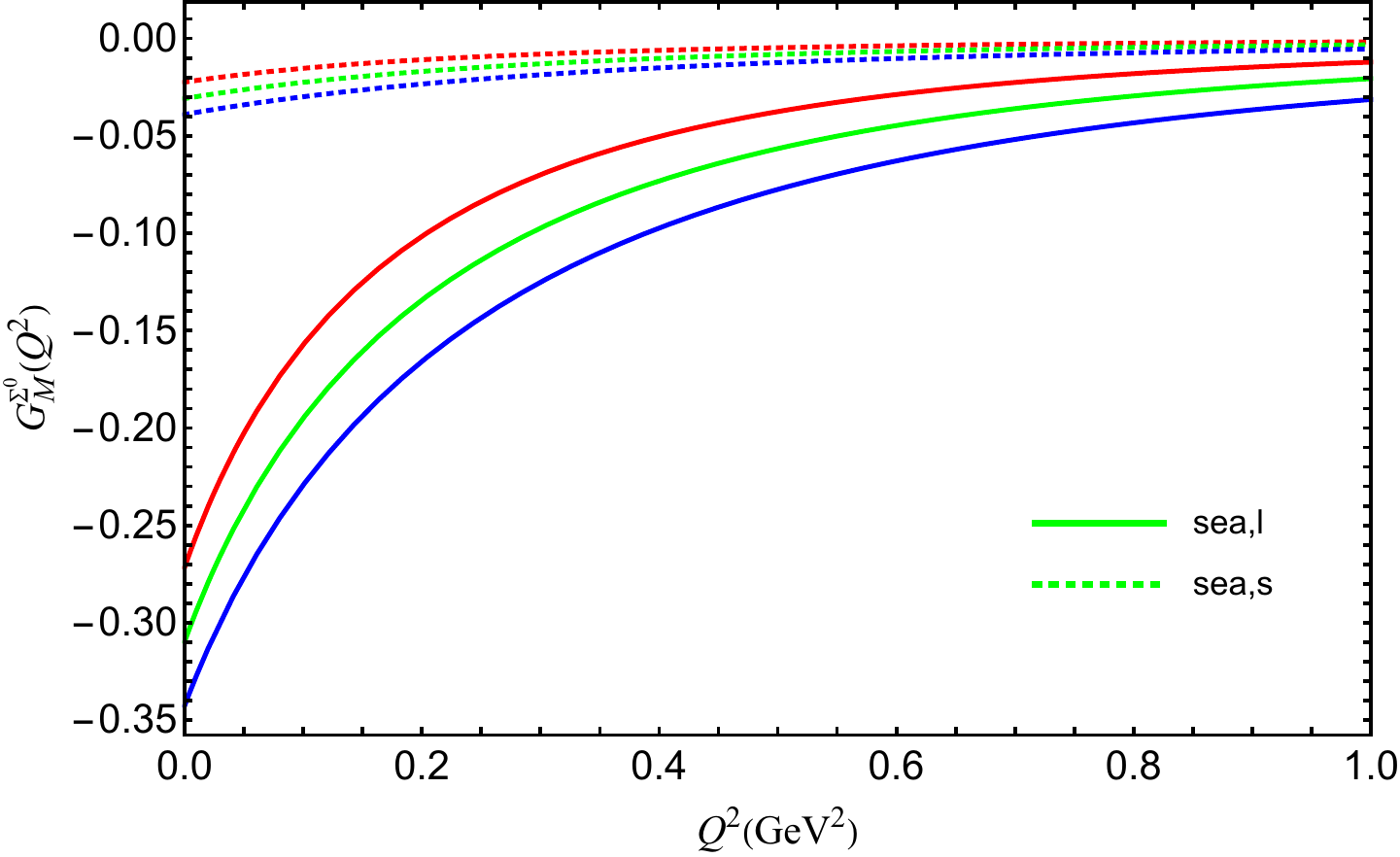}
		\caption{\label{fig:sigma0gm}
			The sea quark contributions to the magnetic form factors of $\Sigma^{0}$ hyperon versus momentum transfer $Q^{2}$. The solid and dashed lines are for the contributions from light sea and $s$ sea, respectively. The lines with color red, green and blue are for $\Lambda=0.8$, 0.9 and 1.0 GeV, respectively.}
	\end{center}
\end{figure}

The sea quark contributions to the electric form factor of $\Sigma^+$ hyperon with the same choice of $\Lambda$, are plotted in Fig.~\ref{fig:Sigma+ge}. 
Because the contributions of quark and anti-quark are from the sea,
their electric form factors are zero at $Q^{2}=0$. 
The additional diagrams are important to get $G_E^{f,\Sigma}(0)=0$. 
They are generated from the additional interaction, which guarantees the local gauge invariance. 
All the electric form factors increase first, and then decrease with the increasing $Q^2$. 
Again, there is a large sea quark asymmetry between $G_E^{u,\Sigma^+}$ and $G_E^{d,\Sigma^+}$. $G_E^{d,\Sigma^+}$ is more than twice as large as $G_E^{u,\Sigma^+}$. 
When $Q^2$ is around \SI{0.3}{\GeV}, $G_E^{d,\Sigma^+}$ is about $0.038 \pm 0.008$.
Compared with the strange electric form factor, $G_E^{d,\Sigma^+}$ is about $10$ times larger. 
Therefore, for both electric and magnetic form factors, 
$G_E^{d,\Sigma^+}$ and $G_M^{d,\Sigma^+}$ are good physical observables for lattice simulation. 
The hyperon form factors would not only shed light on the structure of hyperons, 
but also provide information on nucleon. 
Since the contributions of $u$, $d$ and $s$ quark with unit charge in $\Sigma^-$, 
are the same as $d$, $u$ and $s$ quark contributions in $\Sigma^+$, respectively.
we will not show the results for $G_{E(M)}^{f,\Sigma^-}$ here further.

Now, we discuss the sea quark contributions to the form factors of $\Sigma^0$.
As we have pointed, in nucleon or $\Sigma^+$, there is a large asymmetry between $u$ sea and $d$ sea.
However, in $\Sigma^0$, the sea quark contributions from $u$ and $d$ are the same if we ignore the mass difference between them. 
We can make a direct comparison of the light sea quark form factors between lattice simulation and EFT. 
The sea quark contributions to the magnetic form factor of $\Sigma^{0}$ hyperon versus momentum
transfer $Q^{2}$ are plotted in Fig.~\ref{fig:sigma0gm}. 
From the figure one can see that, 
the sea quark form factors are negative and the shape of the momentum dependence of $G_M^{f,\Sigma^0}$ is close to $G_M^{f,\Sigma^+}$. 
The sea contribution from strange quark in $\Sigma^0$ has similar magnitude as that in nucleon and $\Sigma^+$.  For the light quark contribution, the magnitude of $G_M^{l,\Sigma^0}$ is slightly bigger than $G_M^{u,\Sigma^+}$, and much smaller than $G_M^{d,\Sigma^+}$. 
At $Q^2=0$ with central-valued $\Lambda$, $\mu^{l,\Sigma^0}= -0.310$ and $\mu^{s,\Sigma^0}=-0.031$.

\begin{figure}[H]
	\begin{center}
		\includegraphics[scale=0.6]{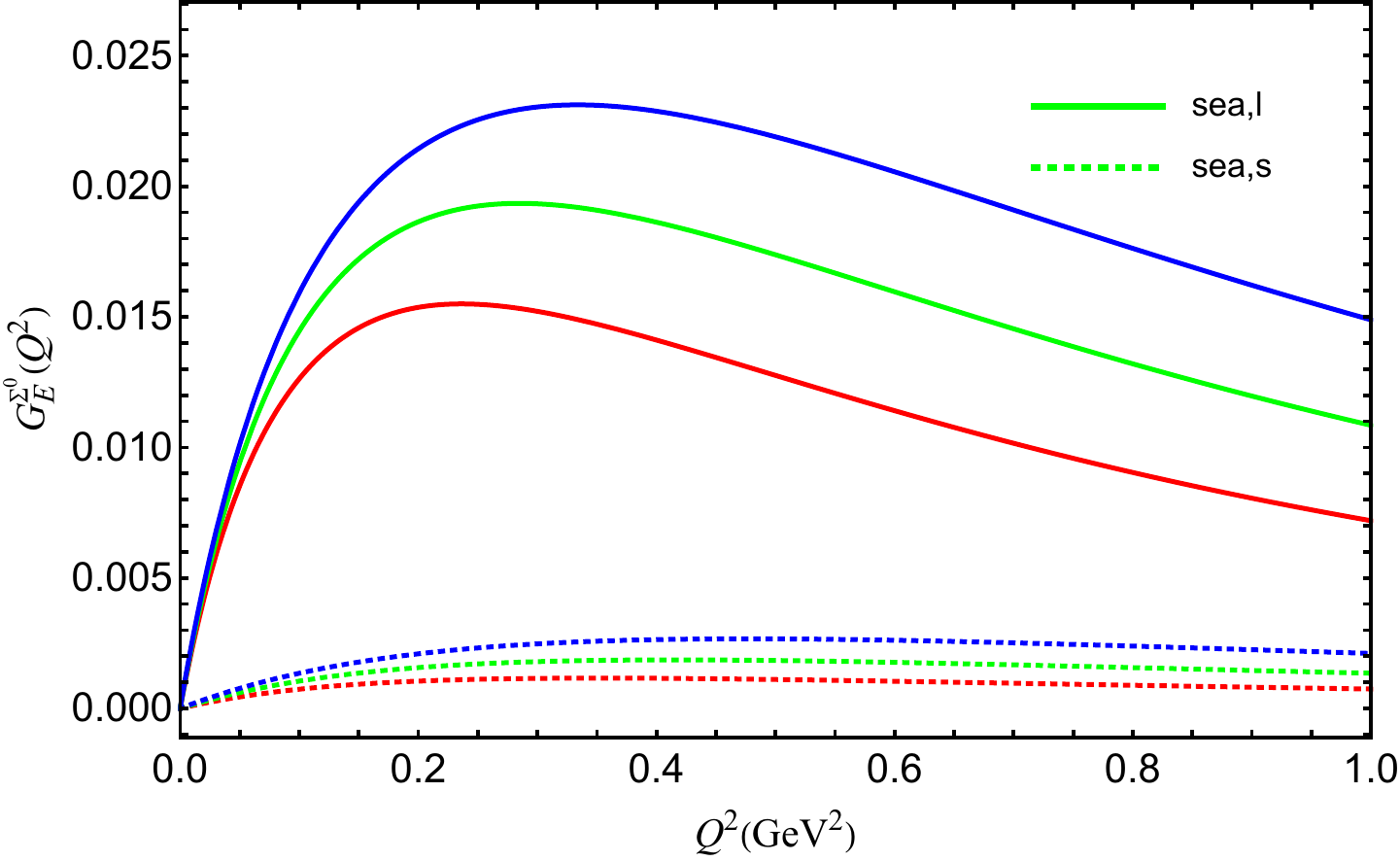}
		\caption{\label{fig:sigma0ge}Same as Fig.~\ref{fig:sigma0gm} but for the electric form factors.}
	\end{center}
\end{figure}

The sea quark contributions to the electric form factors of $\Sigma^{0}$ baryon are plotted in Fig.~\ref{fig:sigma0ge}. 
As for the magnetic case, in $\Sigma^0$, $G_E^{u,\Sigma^0}=G_E^{d,\Sigma^0}=G_E^{l,\Sigma^0}$, and $G_E^{l,\Sigma^0}$ lies between $G_E^{u,\Sigma^+}$ and $G_E^{d,\Sigma^+}$. 
When $Q^2= \SI{0.3}{\GeV^2}$, $G_E^{l,\Sigma^0}\sim 0.019$ with central-valued $\Lambda$. 
For the sea quarks with unit charge, their electric form factors are all positive, 
no matter the total charge of the baryon is $1$, $-1$ or $0$.
Here, the sea quark form factors from the quark-pair, are obtained from the meson loop. 
In the baryon-meson scenario, the baryon is surrounded by the meson where the charge of anti-quark is $-1$. 
The positive electric form factors of sea quark are consistent with this meson loop scenario. 

With the calculated form factors, the contribution to the radii can be obtained as
\begin{equation}
	\langle r_{M(E)}^2\rangle = -6 \frac{{\rm d} G_{M(E)}}{{\rm d}Q^2}|_{Q^2=0},
\end{equation}
where the magnetic radii remain undivided by the corresponding magnetic moments.
Since all the sea quark form factors increase with the increasing $Q^2$ at low momentum transfer, 
the radii are all negative. 
The magnetic and charge radii of $d$--sea in $\Sigma^+$ is larger than those of $u$--sea,
and both of them are more than $10$ times as large as the strange radii.

\section{summary}

\begin{table}[H]
\caption{\label{tab2} The magnetic moments (in units of the nucleon magneton
        $\mu_{N}$), magnetic and charge radii of sea quarks in $\Sigma$ hyperons
        and proton.}
    \centering
    
   \resizebox{\textwidth}{!}{%
        \begin{tabular}{c|c|c|c|c|c|c|c|c|c}
            \hline 
            Baryon & $\mu_{\text{sea}}^{u}$ & $\mu_{\text{sea}}^{d}$ & $\mu_{\text{sea}}^{s}$ & $\langle r_{M}^{2}\rangle_{\text{sea}}^{u}$ & $\langle r_{M}^{2}\rangle_{\text{sea}}^{d}$ & $\langle r_{M}^{2}\rangle_{\text{sea}}^{s}$ & $\langle r_{E}^{2}\rangle_{\text{sea}}^{u}$ & $\langle r_{E}^{2}\rangle_{\text{sea}}^{d}$ & $\langle r_{E}^{2}\rangle_{\text{sea}}^{s}$\tabularnewline
            \hline 
            $\Sigma^{+}$ & $-0.251\pm0.025$ & $-0.676\pm0.083$ & $-0.067\pm0.022$ & $-0.294\pm0.013$ & $-0.811\pm0.029$ & $-0.049\pm0.006$ & $-0.041\pm0.003$ & $-0.119\pm0.007$ & $-0.005\pm0.002$\tabularnewline
            $\Sigma^{0}$ & $-0.310\pm0.035$ & $-0.310\pm0.035$ & $-0.031\pm0.008$ & $-0.394\pm0.015$ & $-0.394\pm0.015$ & $-0.024\pm0.002$ & $-0.061\pm0.003$ & $-0.061\pm0.003$ & $-0.003\pm0.001$\tabularnewline
            $p$ & $-0.111\pm0.005$ & $-0.375\pm0.045$ & $-0.037\pm0.011$ & $-0.142\pm0.011$ & $-0.418\pm0.018$ & $-0.026\pm0.003$ & $-0.036\pm0.002$ & $-0.074\pm0.005$ & $-0.004\pm0.001$\tabularnewline
            \hline 
    \end{tabular}}
\end{table}

The sea quark contributions to the electromagnetic form factors of $\Sigma$ hyperons are
studied within the nonlocal chiral effective theory. 
Both the octet and decuplet intermediate states are included in our calculation. 
The correlation functions in the nonlocal Lagrangian make the loop integrals ultraviolet convergent.
The gauge links guarantee the nonlocal Lagrangians locally gauge invariant.
The expansion of the gauge links generates the additional diagrams and as a result, 
the electric form factors of sea quarks are zero at $Q^2=0$. 
The obtained sea quark magnetic form factors of $u$, $d$ and $s$ are all negative, 
while the electric form factors are all positive. 
They are consistent with the scenario of baryon--meson configurations in dressed baryons.
When $Q^2=0$, $\mu^{d,\Sigma^+}$ ($\mu^{u,\Sigma^-}$) is much larger than the strange magnetic form factor $\mu^{s,N}$. 
For the charge form factors, $G_E^{d,\Sigma^+}$ ($G_E^{u,\Sigma^-}$) is also much larger than $G_E^{s,N}$.
Since there is no valence quark contributions to $G^{d,\Sigma^+}$ ($G^{u,\Sigma^-}$) as $G^{s,N}$,
$G^{d,\Sigma^+}$ ($G^{u,\Sigma^-}$) could be a better physical observables for studying sea quark properties in baryons, for lattice simulations or possible experimental measurements future.
We also found there are large asymmetries of light sea quark form factors in charged $\Sigma$ hyperons. 
For both magnetic and electric form factors, the contributions of $d$ sea are significantly larger than those of $u$ sea. 
To corroborate the comparison of light sea quark form factors between lattice simulation and EFT, 
we calculated the sea quark contributions in $\Sigma^0$. In this case, $G^{u,\Sigma^0}$ equals to $G^{d,\Sigma^0}$. 
The calculations of the sea quark form factors in hyperons,
will not only shed light on the structure of hyperons, but also provide important information on nucleon structure. 
As a summary, we list the magnetic moments and radii of sea quarks in $\Sigma$ hyperons in Table.~\ref{tab2}. 
The corresponding values for proton are also listed for comparison.

\begin{acknowledgments}
	This work was supported by the National Natural Science Foundation of China under Grant No. 11975241.
\end{acknowledgments}


%

\end{document}